\pdfoutput=1 
\documentclass{JINST}
\usepackage{mathrsfs}

\title{Data encoding efficiency in binary strip detector readout}

\author{Maurice Garcia-Sciveres$^b$ and Xinkang Wang$^a$\\
\llap{$^a$} University of California,\\
Berkeley, CA, USA.\\
\llap{$^b$} Lawrence Berkeley National Laboratory,\\
Berkeley, CA, USA.\\
E-mail: \email{mgs@lbl.gov}}

\abstract{A prescription to calculate the minimum number of bits needed for binary strip detector readout is presented. 
This permits a systematic analysis of the readout efficiency relative to this 
theoretical minimum number of bits. Different level efficiencies are defined to include context information and engineering properties needed for reliable transmission, such as DC-balance. A commonly used encoding method is analyzed as an example and found to have an efficiency only of order 50\%. 
A new encoding method called Pattern Overlay Compression is introduced to illustrate how the systematic analysis can guide the construction of more efficient readout methods. 
Pattern Overlay Compression significantly outperforms the above example in the occupancy range of interest.}

\keywords{Particle tracking detectors (Solid-state detectors); Data acquisition concepts; Electronic detector readout concepts (solid-state); Data reduction methods; Information theory}

\begin{document}

\section{Introduction}

Present silicon tracking detectors at the LHC~\cite{LHC} are read out with an external trigger of order 100\,KHz  rate. 
This requires a data bandwidth that was consistent with technology available at the time of construction. As data transmission technology improves, it is natural to consider reading out future tracking detectors with a higher rate. The maximum readout rate achievable depends partly on the efficiency of the data encoding used, and evaluating the limits of readout encoding efficiency is therefore important.  This note develops a general method for analyzing the readout efficiency by considering the theoretical bound of information encoding. The technical details of the method are the central topic of the note. To illustrate the method, a data encoding example for readout of binary silicon systems is analyzed as a function of sensor occupancy. 
A new encoding scheme is then constructed to outperform the original example. 
This demonstrates the power of the analysis method as a tool to make concrete gains, rather than just a theoretical exercise. 
By readout efficiency we mean the number of bits used in practice to extract all the required information from the detector, 
relative to the minimum possible number of bits needed to meet the same requirements. This does not mean the bits used for a single 
event or readout cycle, but the average bits per event for a large ensemble of events.
In practice we define efficiency as the minimum possible number over the actual number of bits, 
so that it has a value between 0 and 1. Clearly our main challenge is to calculate the minimum possible number of bits for given readout requirements.

Our analysis assumes lossless data compression, with the use-case in mind of reading all information from a detector at high rate. An alternative is to read partial information at a high rate, followed by full information at a lower rate. Proposals to implement high rate partial readout followed by low rate full readout have taken the form of a two-level trigger in regions of interest~\cite{roi} or self-seeded high transverse momentum track triggers~\cite{self1,self2,self3}. The present analysis is
specific to silicon strip detectors with binary output. Analysis of pixel detector readout and of readout including signal strength information (as opposed to binary) will involve additional concepts which are the subjects of on-going research, and are beyond the scope of this note.  

To simplify some of the discussion, a 256-channel front end chip with a single serial output is assumed where needed, but 
the results easily generalize to $n$ channels. Typical channel occupancy for strip systems is of order 1\% hit strips per interaction, and so an occupancy range from 0 to 5\% is considered. The roughly 1\% occupancy value is dictated by pattern recognition requirements and applies to any layer radius. In an actual detector the sensor strip dimensions would be adjusted accordingly to achieve everywhere an occupancy in this range, so as far as the readout chip is concerned, every chip in the detector should experience approximately the same input occupancy. 
Note that here occupancy always means average occupancy over a fairly long time (eg. 1 second). The average occupancy is what determines the output data bandwidth requirement, rather than the instantaneous occupancy of a single event. Single event fluctuations can be addressed with local buffering, and we assume that a system will be designed with sufficiently large buffers wherever needed to address single event occupancy fluctuations. 

\begin{figure}[ht]
\centerline{\includegraphics[width=0.5\textwidth]{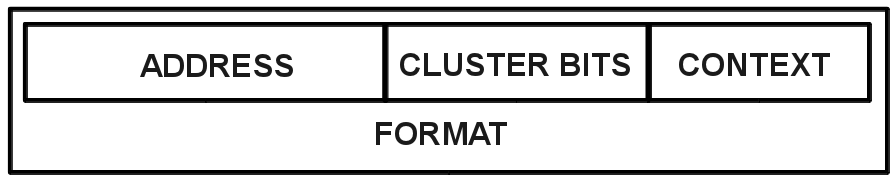}}
\caption{Schematic representation of the four parts that make up the detector output data.}
\label{fig:parts}
\end{figure}

To carry out the analysis we regard the detector output data as composed of four parts shown in Fig.~\ref{fig:parts}. 
The information about which detector strips are hit is contained in the {\em address} and {\em cluster bits} as explained below. 
The {\em context} contains all other information that may be needed for operation or required by the user, while the {\em format} is a kind of wrapper to condition the other parts to meet data transmission requirements, but does not add any new information- all information is contained in the previous three parts. This partitioning allows us to make minimal assumptions needed for the analysis. An actual encoding method need not structure the data in this way for the analysis to be valid. We analyze the readout efficiency in three levels. Level 0 deals only with how efficiently the address part is encoded. Level 1 adds the effect of formatting needed for data transmission. Level 2 adds the final two parts: cluster bits and context. This ordering of the levels goes from most general to most specific. Level 2 would be used to analyze a specific detector, where the details of cluster bits and context are specified. Level 1 can be used to compare detector readout schemes in general, not having to know which detector one is talking about. Finally, Level 0 applies to data compression in general- without even having to assume that it is data from silicon detectors.

\begin{figure}[ht]
\centerline{\includegraphics[width=0.5\textwidth]{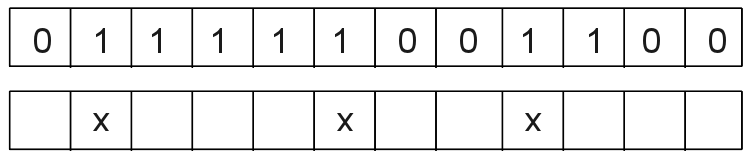}}
\caption{Example bit pattern (top) and resulting addresses (bottom) assuming 2 cluster bits.}
\label{fig:clusters}
\end{figure}

The cluster bits are a characteristic feature of silicon detector readout. They have their origin in the underlying physics of charged particles passing through detector layers. Charged particles cross each layer in approximately random fashion, but each particle may cause a {\em cluster} of one or more adjacent strips to fire. In a binary representation of the hit strips, a cluster is a continuous string of {\em ones}, called a {\em run} in data compression literature. This is commonly exploited in detector applications by encoding the position of only the first strip of a run, followed by a fixed number of {\em cluster bits} to specify the length of the run. Thus the {\em address} identifies a run and the cluster bits specify the length of said run. Note that there is not a 1:1 correspondence between clusters and addresses. The physical process producing a cluster has no limitation on the number of continuous hits, and multiple clusters can also be merged, overlapping one another. On the other hand, a fixed number of cluster bits limit the maximum run that can follow each address. So multiple addresses may be needed to encode a long cluster, or multiple merged clusters may result in a single address. Fig.~\ref{fig:clusters} illustrates how a sample bit pattern results in a collection of addresses for the case of 2 cluster bits (maximum run of 4). 
However, assuming a suitable choice of number of cluster bits will be made, the remaining address part should be 
approximately random. We are thus able to analyze the address encoding using general methods applicable to a random pattern of bits.

The level 0, 1, and 2 efficiencies are derived in the next three sections. The analysis is progressively applied
to the commonly used {\em Channel Address Sparsification} (CAS) encoding scheme, as an illustrative example. The case of multi-chip module readout is then analyzed in Section~\ref{10chip}.
Finally, an example of a new encoding scheme called {\em Pattern Overlay Compression} (POC) is presented in Section~\ref{poc0}.  
Formula derivations as needed are included in the appendix.

\section{Level 0 Efficiency}
\label{sec:compression}

The level 0 efficiency measures how well the address information is compressed by a given encoding scheme. 
We begin by considering the 
theoretical lower bound to the number of bits needed to encode a random bit pattern. 
We then refine the analysis to consider random address patterns and find that the difference is negligible 
in the occupancy range of interest.  
 
Data compression is critical in many fields and is a vast subject of its own~\cite{data-compression}. General algorithms have been developed for lossless compression of large computer files, such as the commonly used zip~\cite{zip}.
A given pattern of bits will be denoted by $x_i$ and the collection of all possible $x_i$ patterns by $X$, 
so that $X= \{ x_1, x_2, x_3,... \}$. 
The smallest average number of bits needed to encode all the patterns in $X$ is given by the 
entropy $H(X)$, where $X$ is regarded as a random variable~\cite{entropy},
\begin{equation}
\label{eq:entropy}
H(X) = -\sum_i p_i log_{_2}(p_i)
\end{equation}
where the sum runs over all possible values of $i$ and $p_i$ is the probability of finding the pattern $x_i$. 
By using the base 2 logarithm the entropy is conveniently expressed in units of bits. 
Note that $H(X)$ is a real number, not an integer, as it is the smallest possible average number bits per pattern and not
the number of bits encoding a specific single pattern in $X$. 
To make this intuitive, consider all $n$-bit patterns as equally likely. The total number of patterns is $2^n$, and so $n$ bits are needed to count them. The entropy should therefore be $n$.  
Equally likely means all patterns occur with the same probability $p_i=1/2^n$. 
Substituting into Eq.~\ref{eq:entropy} indeed yields $H(X)=n$.  
Calculating the entropy for this special case of equally likely states was easily done by simply counting states, 
without resorting to Eq.~\ref{eq:entropy}, 
but for an arbitrary probability distribution simple counting will not be adequate, while Eq.~\ref{eq:entropy} applies in general. 

\begin{figure}[tb]
\centerline{\includegraphics[width=0.65\textwidth]{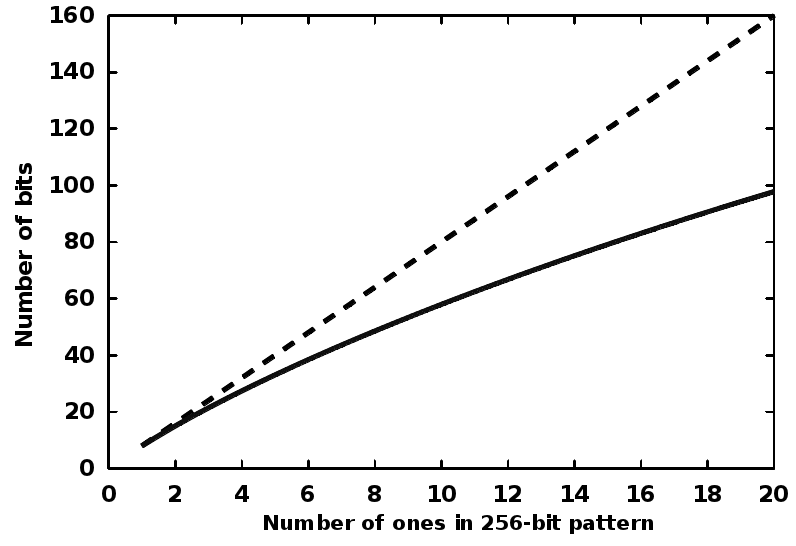}}
\caption{The entropy, or minimum average number of bits needed to count the total number of 256-bit patterns containing a given number of {\em ones} (solid). 
The dashed line, equal to 8 times the number of ones, is shown for comparison.}
\label{fig:logsk}
\end{figure}

Since the variable of interest for detector readout is occupancy, 
let us consider the number of $n$-bit patterns with a given occupancy, $X^n_k$, where every pattern has exactly $k$ {\em ones}. 
For the case of completely random occupancy the total number of possible patterns with occupancy $k$ can again be simply counted, 
and it is nothing other than the binomial coefficient $n \choose k$.
The entropy should be the number of bits needed to count the patterns, $log_{_2}$$n \choose k$. 
The same result is obtained by using Eq.~\ref{eq:entropy} for $n \choose k$ patterns each occurring with equal probability $1/$$n \choose k$.
Fig.~\ref{fig:logsk} shows the entropy $H(X^n_k)$ as a function of $k$ for $n=256$.
A dashed line of slope $8k$ is shown for comparison, which corresponds to the commonly used scheme of listing the addresses
of all non-zero channels. We will refer to this as Channel Address Sparsification (CAS). 
Note that 5\% occupancy, which is the upper limit we are considering, corresponds to just under 13 {\em ones} in this figure. 
Full frame readout (one bit per strip whether hit or not) would use 256 bits regardless of occupancy.

Rather than the entropy of a random bit pattern, we wish to calculate the entropy of a random address pattern. The difference is that certain address combinations are excluded by the use of cluster bits. For example, the hit pattern ...0110... is perfectly valid, but the address pattern ...0XX0... is not. When using cluster bits,
 the hit pattern ...0110... would result in the single address ...0X00...
This restriction can be expressed in general by the condition that addresses can not be contiguous; 
all addresses must be separated by at least one strip. 
This condition leads to a modified expression for the entropy. 
While the entropy of $k$ random {\em ones} was $log_{_2}$$n \choose k$, the entropy of $k$ random addresses is 
$log_{_2}$$n-k+1 \choose k$ (see Appendix for derivation). 
The difference between the expressions grows with increasing $k$ (at $k=1$ they are identical). 
However, for $k \ll n$ the difference is negligible: at $k/n=$5\% the difference is only 1.3\%.
The remainder of this note will consider {\em address occupancy} rather than raw occupancy. Note that if the addresses are not 
exactly random then the Eq.~\ref{eq:entropy} entropy will be smaller, so the randomness assumption makes our entropy bounds conservative.

\begin{figure}[htb]
\centerline{\includegraphics[width=0.65\textwidth]{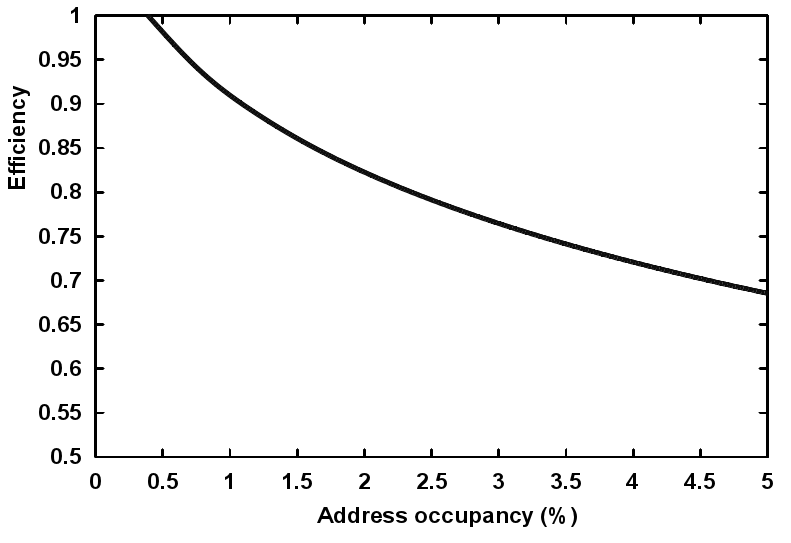}}
\caption{Address encoding efficiency for channel address sparsification (CAS) as a function of address occupancy in percent, 
for a 256-bit pattern.}
\label{fig:effic1}
\end{figure}

\begin{figure}[htb]
\centerline{\includegraphics[width=0.65\textwidth]{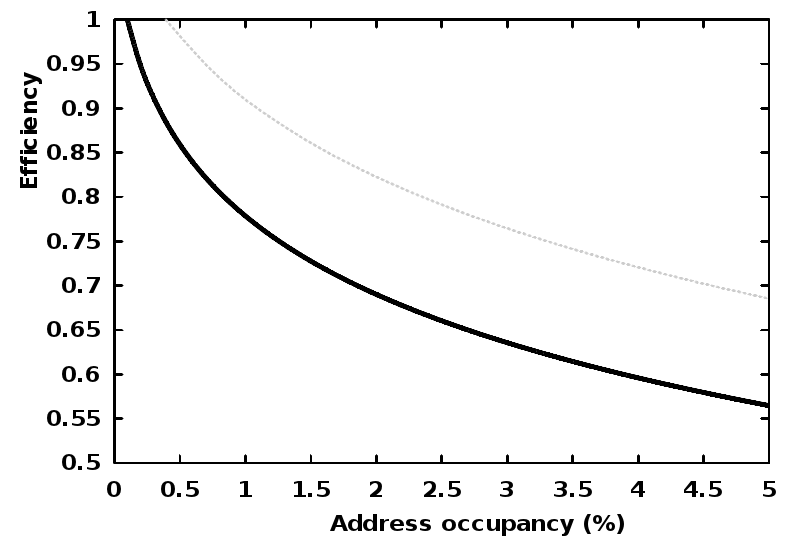}}
\caption{Address encoding efficiency for channel address sparsification (CAS) as a function of address occupancy in percent, 
for a 1024-bit pattern. The dotted line is the efficiency curve for a 256-bit pattern, for comparison. }
\label{fig:effic2}
\end{figure}

As mentioned in the introduction, 
we define the encoding efficiency as the ratio of the entropy over the number of bits, $B$, used by any given encoding method
(the entropy being the smallest possible number of bits needed to encode the information). 
\begin{equation}
\label{eq:efficiency0}
\epsilon_0 = H/B
\end{equation}
where we have used the symbol $\epsilon_0$ to denote the level 0 efficiency. 
This efficiency is shown for CAS encoding in Fig.~\ref{fig:effic1}.
Fig.~\ref{fig:effic1} is approximately the ratio of the solid line over the dashed line in Fig.~\ref{fig:logsk}, with the 
difference that we consider address occupancy rather then raw occupancy, as explained above.   

In Fig.~\ref{fig:effic1} one can see that the efficiency is $>$90\% for 1\% occupancy, which seems very good, 
yet CAS encoding is not generally used for data compression of sparse bitstreams in commercial applications. 
Other methods, such as Huffman codes and prefix compression are used instead~\cite{data-compression}. 
One may wonder why. The reason is that CAS is only efficient for compressing a short bitstream of very low occupancy. 
In general applications, bitstreams are much longer (megabits) and occupancy is not tightly constrained to very low values. 
As $n$ increases in $X^n_k$, the fractional occupancy ($k/n$) at which CAS works efficiently drops. 
Fig.~\ref{fig:effic2}
shows the CAS efficiency vs. address occupancy for $n=1024$. 
The efficiency is lower 
because CAS encoding is only efficient for patterns where $k$ is small in absolute terms.
Thus, for any given fractional occupancy $k/n$, the efficiency of CAS encoding can be made arbitrarily low 
by choosing larger and larger $n$. 
So even though CAS may be an efficient choice for compressing a single hit pattern from a single chip, 
it is not efficient when considering the aggregated data from a significant number of chips or patterns. 
The efficiency of CAS as a function of $n$ 
and fractional occupancy $\alpha \equiv (k/n)$ is given in Eq.~\ref{eq:casefficiency} (see Appendix for derivation).  
This shows how the CAS efficiency drops with increasing $n$. 
\begin{equation}
\label{eq:casefficiency}
\epsilon_0(n,\alpha) \approx {1-ln(\alpha) \over ln(n)}
\end{equation}  
  
\subsection{Choice of $n$ and Entropy per Address}

Eq.~\ref{eq:casefficiency} suggests that the efficiency of a given encoding method will depend on the choice of $n$. 
For detector readout applications there are several choices given by the hardware design. Most obviously there is the 
number of channels in a single readout IC, but it is a relatively small value: historically 128 and more currently 256. 
On the other hand,
the number of channels in an entire detector will be in the millions. The readout must allow the unique location of each 
hit within the detector.
Therefore, a natural choice of $n$ from an encoding efficiency standpoint is the number
of channels served by a single data link, denoted by $n_L$. As a group, the position of these $n_L$ channels in the detector 
can be established by ``following the readout cable''. 
Therefore, the information in the data stream on said ``cable'' need only uniquely identify each 
channel within the group, not the absolute location in the detector. Of course, a data link need not 
be a physical cable, but rather a part of the detector that is permanently mapped to a data stream in the data acquisition system
by hardware connections and/or configuration. In typical silicon strip detectors this corresponds to a {\em module}. A module typically 
contains multiple readout ICs and a number of channels of order 1000. 

\begin{figure}[htb]
\centerline{\includegraphics[width=0.65\textwidth]{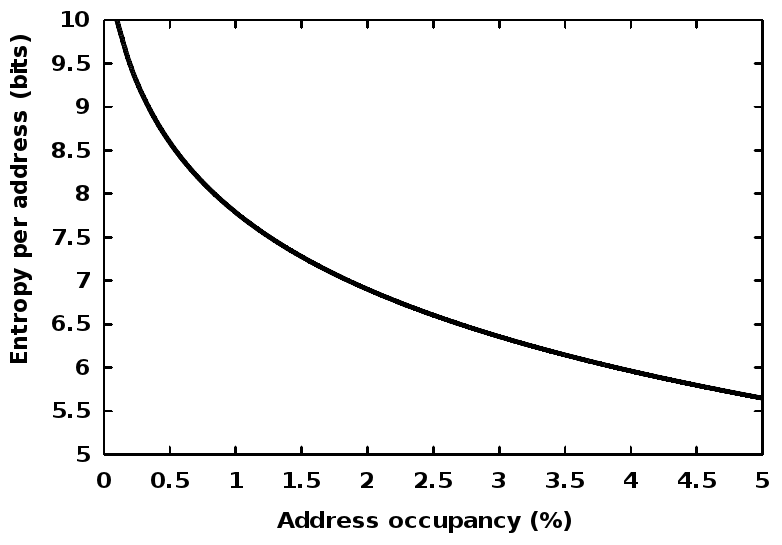}}
\caption{Entropy per address vs. address occupancy for a 1024-bit pattern.}
\label{fig:entropycluster}
\end{figure}

As we have seen above, CAS encoding is more efficient for $n=256$ than for $n=1024$.
Use cases of CAS encoding in past detectors in fact relied on a 7-bit address space (corresponding to 128-channel ICs), 
and identified different ICs within a module using a few-bit IC address space. This is a form of the Prefix Compression
method~\cite{data-compression}. For the case of $n_L=1024$, 
patterns could be encoded with a 3-bit IC identifier and a 7-bit subaddress. The reason this is in principle more efficient than a plain 10-bit address space is that the IC identifier is only transmitted 
once for all addresses within that IC, instead of needing a full 10 bits for each address. 
As long as the occupancy is high enough that multiple
addresses within the same IC are common, splitting the 10-bit address space into an x-bit prefix and a y-bit subaddress seems advantageous. 
However, a new problem arises: one must encode how many addresses there are in each IC 
(or alternatively when the address words from one IC end and those of the next IC begin). 
Thus, the efficiency gain is not as large as naively expected. 
Nevertheless, this suggests looking to the entropy per address, $H^n_k/k$, for guidance of how many bits might be ideal to use for subaddress encoding vs. prefix. 
Fig.~\ref{fig:entropycluster} shows the entropy per address as a function of address occupancy for $n_L = 1024$.  
This shows that at 1\% occupancy the entropy content is about 8 bits per address. 
Therefore, a Prefix Encoding method should use fewer than 8 bits for the subaddress
to have any hope of approaching the entropy limit.

\section{Level 1 Efficiency}
\label{sec:engineering}

So far the analysis has been abstract in asking how a pattern of bits can be compressed, without worrying about how the bits may be stored or transmitted. 
These practical aspects place important constraints on the format of the encoded data, which in turn affects the achievable compression. For reliable transmission a bit pattern with ``engineering properties'' is needed. These include 
DC-balance, framing, and error detection with or without error correction. 

DC-balance means that up to a reasonably short time interval an arbitrarily chosen segment of data should contain equal numbers of {\em ones} and {\em zeroes}. Different implementation methods use different length intervals over which balance is achieved. The effect of balancing is to constrain to a narrow band the frequency spectrum of a serial bit stream viewed as an A/C signal. This permits optimal functioning of driver and receiver circuits as well as transmission lines. 
Typically the same hardware can transmit a significantly higher bit rate without errors when DC balance is used than when it is not,
allowing more information to be transmitted per unit time even if extra bits are needed to achieve DC balance.  

Framing allows the receiver of a serial bit stream to determine where meaningful bit patterns begin and end, while error detection allows the receiver to know when a transmission error has occurred. One can simply know that an error occurred and therefore discard the affected data, or one can have enough information to undo the error and correct the data. The level of error detection and correction varies widely depending on what the application requires. 
For silicon detector hit data error detection alone is appropriate, 
since no detector is 100\% efficient to begin with, 
and therefore loss of a small amount of data during transmission is acceptable (typical hit inefficiencies not related to data transmission are of order 1\%). 

Several commonly used  {\em transforms} can be applied to an arbitrary {\em raw} bit pattern to produce 
a {\em transformed} pattern with engineering properties. The inverse transform can then be applied to the 
transformed pattern in order to recover the original raw pattern. The 8b/10b (8 bit 10 bit)~\cite{8b10b} transform divides the 
raw pattern into 8-bit words or {\em symbols} and replaces each one with a 10-bit symbol. 
The 10-bit symbols should ideally be chosen to always have 5 of each {\em ones} and {\em zeroes},
but since there are only $10 \choose 5$ $=252$ such symbols 
(while there are 256 possible 8-bit symbols), 10-bit symbols with 6 or 4 {\em ones} are also used, and 
8-bit symbols can be mapped to more than one 10-bit symbol. DC balance is guaranteed over a 20-bit span by alternating the use of 
symbols with 6 or 4 {\em ones}.  A greater number than 256 10-bit symbols is used, effectively allowing 
the transformed pattern to contain extra information that was not present in the raw pattern, which the user can exploit in
different ways, for example to mark start and end of transmission. 
Framing is provided by the fact that the number of valid symbols is much less that the total number (1024) of 
symbols with 10 bits, so the frame alignment is varied until ``only'' valid symbols are seen.
Quotes were used for ``only'' because invalid symbols are still tolerated at a low rate without triggering frame
adjustment. Such infrequent invalid symbols are interpreted as transmission errors. 
In fact, flipping one bit in any of the valid 10-bit symbols results in an invalid symbol. 
The 8b/10b transform is simple to implement in hardware and clearly suited to raw data consisting of 8-bit words.

Scrambling transforms calculate each successive bit of the transformed pattern as a function 
of previous bits in the transformed pattern itself as well as the incoming bits in the raw pattern.
A descrambling algorithm reverses the operation to recover the raw pattern. 
This type of transform provides DC balance {\em only}, without increasing the length of the bit pattern. 
One must then provide for framing and error detection separately. 
Ethernet connections use the 64b/66b transform~\cite{6466},
which as the name suggests uses 66 bits to transmit 64-bit content. Unlike 8b/10b, 64b/66b consists of a scrambler plus 
a 2-bit header, which provides framing, the ability to switch between two encoding formats, 
and error detection for the header only, not the scrambled data.  
Another common method for framing a random bit pattern is to insert 
fixed {\em synchronization} bits with a periodic interval. The receiver can then scan for such a repeating pattern.
Similarly, error detection can be implemented by inserting extra bits that are a function of the raw pattern. The simplest 
incarnation is the parity bit. Hamming coding~\cite{hamming} provides a higher level of detection and even correction at the 
expense of more added bits. 

\subsection{Efficiency with Engineering Properties}
\label{EEP}

\begin{figure}[htb]
\centerline{\includegraphics[width=0.65\textwidth]{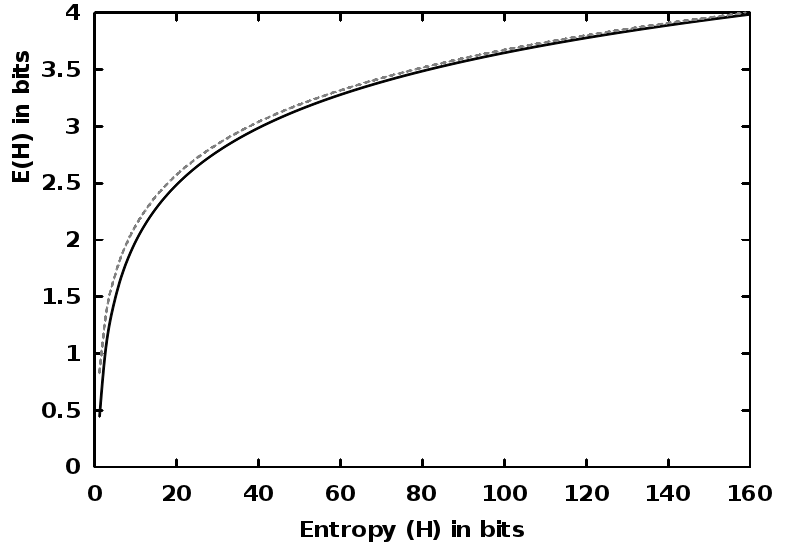}}
\caption{Overhead $E(H)$ to account for engineering properties needed for data transmission.  
Analytic approximation (solid) and numerical calculation (dashed).}
\label{fig:hm}
\end{figure} 

It is clear from the above examples that adding engineering properties to a raw bit pattern entails increasing the number of bits. 
Therefore, the level 0 efficiency seems ``unfair'' when considering data transmission, 
since it would be impossible to achieve 100\% efficiency. 
The level 1 efficiency, $\epsilon_1$, takes into account the overhead due to engineering properties. 
This is accomplished by making the substitution 
$H \rightarrow H + E(H)$, leading to
\begin{equation}
\label{eq:efficiency1}
\epsilon_1 = {H+E(H) \over B_1}
\end{equation}
where $B_1$ is the number of bits used by a given encoding method including engineering properties.
Note that we anticipated that the number of bits one needs to add should depend on $H$. 
Unfortunately, we have no absolute measure of the smallest possible $E(H)$ needed to achieve DC-balance, framing, and error detection. Nevertheless we derive a plausible expression for $E(H)$. We do this by generalizing the 
8b/10b transform concept. We consider a pattern of $H$ bits as an $H$-bit symbol. There are $2^H$ such 
symbols (note that this expression is valid for non-integer $H$ even though the concept of $H$ bits is not). 
We therefore want to find the value $m$ such that $2m \choose m$$\ge 2^H$. $2m \choose m$ is the number of $2m$-bit
symbols with exactly $m$ {\em ones}, and thus DC-balanced by construction. As with 8b/10b, by using only DC-balanced symbols 
one achieves not just DC-balance, but also framing and error detection, which are all the things we want to 
include in the level 1 efficiency. (As noted before, 8b/10b performs
a little better by alternating not perfectly DC-balanced symbols, but we will ignore this). 
We can turn the expression around 
and for every integer $m$ calculate the entropy content of all the DC-balanced $2m$-bit symbols,
\begin{equation}
\label{eq:h2m}
H_{2m} = log_2 {2m \choose m}
\end{equation}
For these special, discrete entropy values, $E(H_{2m}) = 2m - H_{2m}$. 
Eq.~\ref{eq:EH} gives an approximate analytic expression for a continuous range of $H$ (derived in the Appendix). 
Fig.~\ref{fig:hm} plots this expression (solid) compared to a numerical calculation (dotted). 
\begin{equation}
\label{eq:EH}
E(H) \approx {log_2(\pi H) - 1 \over 2}
\end{equation}

\section{Level 2 Efficiency}
\label{sec:context}

The level 2 efficiency includes all the elements shown in Fig.~\ref{fig:parts}. However, the number of cluster bits 
and the amount of context information needed per transmission are detector and experiment-specific choices.
We therefore will not consider how to determine optimum number of bits, or how to best encode context information.
Instead, we assume values $C$ and $\nu$ for the number of context and cluster bits, respectively. These lead to a 
fixed overhead in the level 2 efficiency, which is considered equal for any encoding method.  
We thus arrive at a level 2 efficiency definition,
\begin{equation}
\label{eq:efficiency2}
\epsilon_2 = {H+E(H)+C+\nu K \over B_2 },
\end{equation}
where $\nu$ is the number of cluster bits per address used to specify run length and $K$ is the number of addresses used to encode the
raw pattern. 
$B_2$ is the number of bits used by a particular encoding method, including context and cluster bits. 
We did not explicitly break out $C$ and $\nu K$ in the denominator, because the way they fold into the total number of bits may depend 
on the encoding method used. 

\section{Case Study of 10-Chip Module with Fixed Packet Format}
\label{10chip}

\begin{figure}[ht]
\centerline{\includegraphics[width=0.65\textwidth]{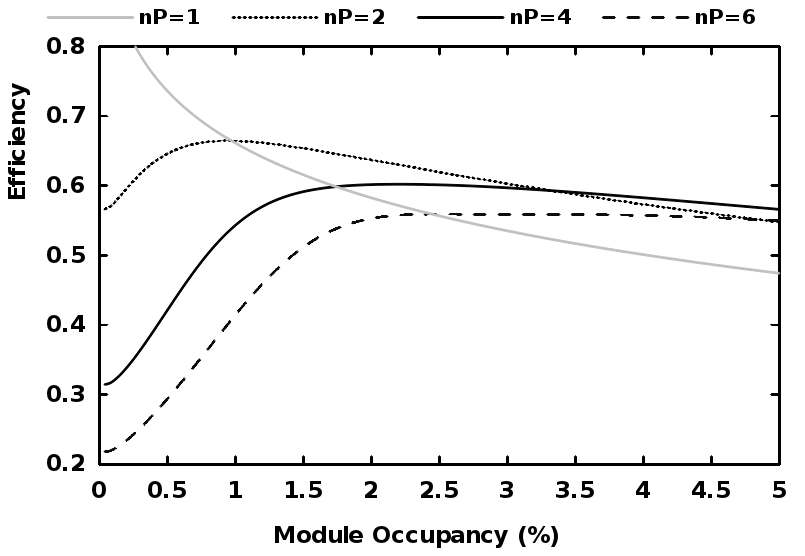}}
\caption{Level 0 average efficiency for fixed size packet CAS encoding in a 10-chip module for several choices of packet size. A packet size of 1 is equivalent to plain CAS encoding with 12 bits per cluster. The address occupancy is for the whole module.}
\label{fig:packet}
\end{figure}

We study the case of encoding the hit data from a 10-chip module with 256 channels per chip, corresponding to the silicon 
strip detector readout proposed in \cite{sct}. We consider the hit data encoded in fixed length packets, each packet containing a chip ID and a fixed number of 8-bit addresses denoted by $n_P$.
As already discussed the cluster bits can be separated out and are not considered in $\epsilon_0$ of Eq.~\ref{eq:efficiency0}. 
Thus at level 0 the packet length is $4+8n_P$, where 4 bits are needed to 
count 10 chips. We study this efficiency as a function of address occupancy given by $K/2560$, 
where $K$ is the total number of addresses needed for the full module. 
The $i^{~th}$ chip  will use $k_i$ addresses and $K = \sum k_i$. We calculate the average efficiency $<\epsilon_0>$ assuming the 
$K$ addresses are randomly distributed among the 10 chips. 
The results are shown for several values of $n_P$ in Fig.~\ref{fig:packet}. 
It is interesting to note that $\epsilon_0$ has a plateau over a range of occupancy depending on $n_P$, 
which seems like an appealing property. 
However, the value of the plateau is only in the neighborhood of 60\% efficiency, without a strong dependence on $n_P$. 

\begin{figure}[hb]
\centerline{\includegraphics[width=0.65\textwidth]{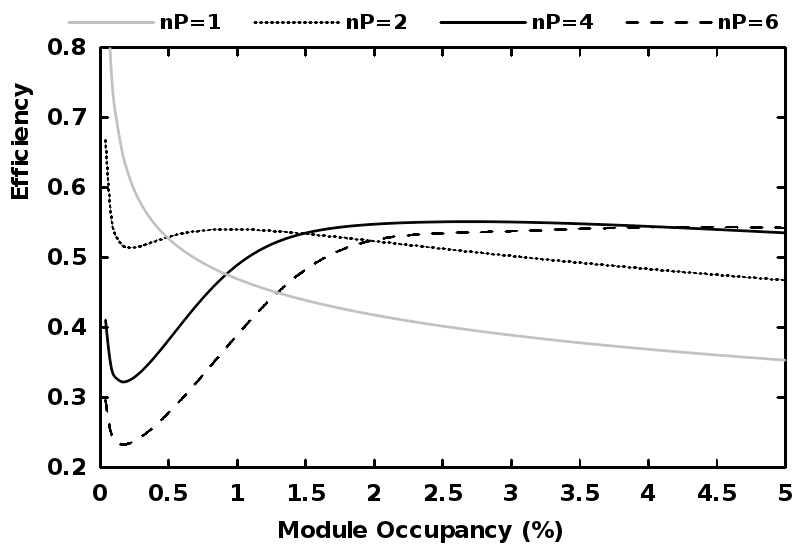}}
\caption{Level 2 average efficiency for fixed size packet CAS encoding in a 10-chip module for several choices of packet size.
Each packet includes 6 context bits and 2 cluster bits per address. The address occupancy is for the whole module.}
\label{fig:packet2}
\end{figure} 

We next consider $\epsilon_2$ of Eq.~\ref{eq:efficiency2}, 
which includes context information, cluster bits, and engineering properties. 
We arbitrarily take $v=2$ cluster bits and $C=6$ context bits for the sake of illustration. 
For engineering properties we make the minimal assumption of scrambling plus 2 bits per packet, one for framing and the other a parity bit.
Thus,
\begin{equation}
\epsilon_2 = { H + E(H) + 6 + 2K \over P [ 2 + 4 + (8+2)n_P + 6] }
\end{equation}
where $P$ is the average number of packets used for transmission and the terms in the denominator correspond to: framing and parity (2), chip number (4), address and cluster bits ($(8+2)n_P$), and finally context bits (6).  
The results are shown in Fig.~\ref{fig:packet2}. 
Larger packets (i.e. larger $n_P$) now perform better at large module occupancy, not surprisingly, as the context and engineering property bits 
must be added to every packet. The efficiency plateaus flatten and move to higher occupancy, and in all cases the efficiency for
reasonable occupancy is 55\% or less. This suggests that a significant gain is possible relative to this encoding method.

\section{Pattern Overlay Compression} 
\label{poc0}

Understanding the sources of inefficiency using the method we have developed enables the construction of better encoding methods. 
As an example we develop a new encoding method called Pattern Overlay Compression (POC),
tailored to the specific case of multi-chip readout. 
This method is based on the observation that an $N$-bit pattern with $N/2$ ones has entropy that approaches $N$ for large $N$
(consider the approximation $2x \choose x$$ = 4^x/\sqrt{\pi x}$). Thus, a long pattern with 50\% random occupancy will have 
high efficiency. 
One can therefore try to represent the hit data using high occupancy patterns, rather than compressing low occupancy ones. 
The POC method constructs one high occupancy pattern by overlaying multiple low occupancy patterns of equal length.
In order for this compression method to be lossless, the low occupancy source patterns must be labeled and the labels attached
to the resulting high occupancy pattern. The method is best explained algorithmically as follows. 

Let the source patterns all have $N$ bits. The result pattern will consist of $N$ {\em zeroes} and $K$ {\em ones}, 
where $K= \sum k_i$ is the total number of {\em ones} in the source patterns. 
Thus the result pattern has a total of $N+K$ bits. The {\em zeroes} in the result pattern act as delimiters, effectively 
creating a table with $N$ bins. In each bin of this table, a number of {\em ones} is placed, 
which act as flags to indicate how many source patterns
had a {\em one} in this location: 3 flag {\em ones} means 3 source patterns and so on. Table~\ref{tab:POC} illustrates
an example with 4 source patterns with $N=8$. 
The results pattern from Table~\ref{tab:POC} is 10110001100010, where there is one {\em zero} for every vertical line in the
table, except the first one, which we have chosen to ignore as it is not needed. The number of flag {\em ones} in the result pattern 
(six in this case), is the total number of {\em ones} from all the source patterns.

The result pattern alone is not enough for lossless compression. We also need to label which source pattern each flag refers to. 
In our example from Table~\ref{tab:POC}, 
the result pattern should be followed by DACBCB, to indicate which source pattern each of the 
six flags belongs to. Alternatively, a label can be placed right after each flag instead of all the labels at the end. 
Thus, if there are $n_s$ source patterns, the result pattern must include or be followed by $K.log_2(n_s)$ bits. 
The full POC bitstream for our example is therefore: 
10110001100010(11)(00)(10)(01)(10)(01) with labels at the end, or 1(11)01(00)1(10)0001(01)1(10)0001(01)0 
with a label after each flag, 
where the parentheses are just a visual aid to highlight the pattern labels. 
Further compression of the pattern labels is possible whenever there are multiple labels in the same bin. 
If the labels are always written monotonically (smallest to largest) then the most significant bits of multiple labels in the same bin may be known from the value of the first label. 
In the example given, the pair of labels (01)(10) could be abbreviated as (01)(0) without loss of information, since from the first label and the condition that labels are written in increasing order it is already known that the first bit of the second label must be a {\em one}.

\begin{table}[h]
\begin{center}
{\setlength{\tabcolsep}{1pt}
\begin{tabular}{r|c|cl|cl|cl|cl|cl|cl|cl|}
\hline
Source pattern A:~~&~~~0~~~&~~&1&~~&0&~~&0&~~&0&~~&0&~~&0&~~&0\\
\hline
Source pattern B:~~&~~~0~~~&&0&&0&&0&&1&&0&&0&&1\\ 
\hline
Source pattern C:~~&~~~0~~~&&1&&0&&0&&1&&0&&0&&0\\
\hline
Source pattern D:~~&~~~1~~~&&0&&0&&0&&0&&0&&0&&0\\
\hline\hline
Result pattern:~~  &~~~1~~~&~~&11~~&~~~&~~~~~&~~~&~~~~~&~~~&11~~&~~~&~~~~~&~~~&~~~~~&~~&1\\
\hline
\multicolumn{1}{r}{} & \multicolumn{2}{c}{~~~~~~0} & \multicolumn{2}{c}{~~~~0} & \multicolumn{2}{c}{~~0} & \multicolumn{2}{c}{~~0} & \multicolumn{2}{c}{~~~~0} & \multicolumn{2}{c}{~~0} & \multicolumn{2}{c}{~~0} & \multicolumn{1}{c}{~~~0} \\
\end{tabular}}
\caption{\label{tab:POC}
Example to illustrate the calculation of the result pattern in POC. The {\em zeroes} below the table indicate which bin boundaries (vertical lines) are represented by a {\em zero} in the result pattern. This result pattern would be written 
as 10110001100010.}
\end{center}
\end{table} 

\begin{figure}[h]
\centerline{\includegraphics[width=0.65\textwidth]{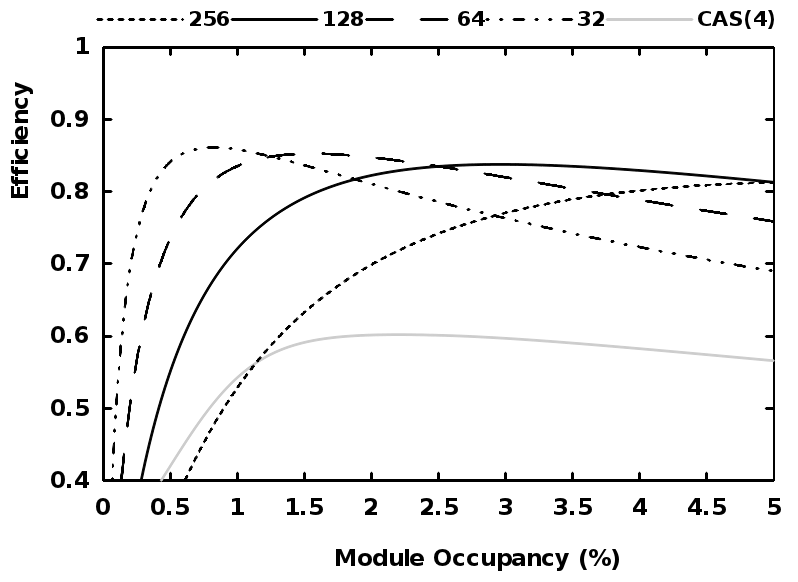}}
\caption{Level 0 efficiency for POC encoding in a 10-chip module for different result pattern number of bins (256, 128, 64, 32),
as a function of module address occupancy. 
Also shown for comparison is the average efficiency of fixed packet CAS encoding with $n_P$=4.}
\label{fig:poc1}
\end{figure}   

If no compression is applied to multiple labels in the same bin, 
the total number of bits used by POC is simple to calculate. For the case of a 10-chip module with address occupancy $K$, 
chip-wise POC encoding (each chip is one source pattern) would lead to $B_0=256+K+4K$ bits, 
as there are 256 bins and 4 bits are needed to label the 10 chips. 
Since the result pattern is a representation of a table where each {\em zero} marks one bin, it is always possible to re-bin this table and so change the number of {\em zeroes}. 
For example, instead of 256 bins we can use 128 bins, where each new bin combines two of the original bins. An additional label bit will then be needed to identify which original bin was occupied. Thus, $B_0=128+K+(4+1)K$ bits, $B_0=64+K+(4+2)K$ bits, etc, would also be valid and practical POC encoding lengths.
 
Fig.~\ref{fig:poc1} shows $\epsilon_0$ for 
10-chip module POC encoding for various choices of result pattern binning, between 256 and 32. No compression has been applied to multiple labels in the same bin and $n_P=4$ fixed packet $\epsilon_0$ is  also shown for comparison.

\subsection{Practical Application to Interconnected Chips}

\begin{figure}[h]
\centerline{\includegraphics[width=0.65\textwidth]{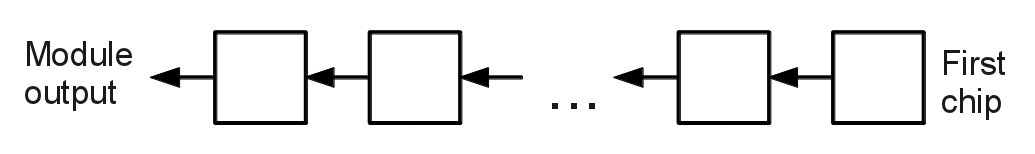}}
\caption{Diagram illustrating a number of chips connected in a daisy chain configuration. The arrows indicate data flow direction.
The number of chips is arbitrary.}
\label{fig:chain}
\end{figure}   

POC is naturally suited to readout of multiple chips that are daisy-chained together, as shown in Fig.~\ref{fig:chain}. 
The format is valid for any number of chips, even a single chip. The first chip in the daisy chain sends out its data
in POC format. The next chip receives this data, adds its bin contents, and sends the combined data to the next chip. 
This is repeated down the chain until the last chip, which produces the module output with the data from all chips. 

\subsection{Level 1 and Level 2 Efficiency of POC}

For addition of engineering properties it is possible to make the same minimal assumptions of scrambling, framing, and use of parity bits that were made for analyzing a fixed packet format. However, the presence of a predetermined number of {\em zeros} in every result pattern 
presents an opportunity to develop a special method unique to POC, so we propose one such possible method here. 
This method is slightly less efficient than the minimal assumptions made for fixed packet format and therefore makes the example more conservative. 
In analogy to 8b/10b, we use a counter to keep track of DC-balance. The counter is incremented for 
every output {\em one} and decremented for every {\em zero}. 
The counter reports whether the output is so far perfectly balanced, has too many {\em ones} or too many {\em zeroes}. 
The proposed encoding starts from the result pattern with embedded 
source pattern labels. Each original {\em zero} (bin delimiter) remains a {\em zero} if the DC-balance counter is perfectly balanced 
or has too many {\em ones}, but is replaced with a {\em one} otherwise. Thus, for example, the empty result pattern 000000... 
will become 010101... The bin content flags, which were always {\em one} in Table~\ref{tab:POC}, remain {\em one} if the counter is balanced or has too many {\em ones}, 
but are replaced with {\em zero} otherwise. 
So far it can be said that bin delimiters act to restore DC-balance, while bin content flags act to increase DC-imbalance. 
The ideal effect of a label should therefore be to restore DC balance. 
As each flag is followed by a label, the combination of flag plus label would then be DC-balanced. 
For the example of a 10-chip module, one can use 5-bit labels containing exactly 3 {\em zeroes} and 2 {\em ones}. 
There are exactly 10 such labels. Furthermore, the bit-wise inverse of each label contains 3 {\em ones} and 2 {\em zeroes}. 
If the DC-balance counter is balanced or has too many {\em ones}, a label with 3 {\em zeroes} is used, otherwise its 
bit-wise inverse (which has 3 {\em ones}) is used. 
Thus the labels act to restore DC-balance. Table~\ref{tab:pocdc} summarizes this encoding. Note that this is a special case for this example, but illustrates a method that can be generalized. Note also that 5-bit symbols are used as labels where only 4-bit codes would be needed without DC-balance, which is the same overhead as in 8b/10b.

\begin{table}[h]
\begin{center}
\begin{tabular}{|l|c|c|c|}
\hline
DC-balance counter & balanced & too many {\em ones} & too many {\em zeroes} \\
\hline
Bin delimiter      &  0       &    0                &     1 \\
\hline
Content flag                &  1       &    1                &     0 \\
\hline
Pattern label      & 00011    &  00011              & 11100 \\
\hline
\end{tabular}
\caption{\label{tab:pocdc}
Illustration of the method to add engineering properties to the POC result pattern. The pattern label shown is representative 
to indicate the number of {\em zeros} and {\em ones} in the label, but there are 10 possible such labels (not shown).}
\end{center}
\end{table} 

The result pattern DC-balanced this way contains the needed labels, but context information and cluster bits must still be added. 
For simplicity of this estimate, we propose to add such information immediately following the result pattern using 8b/10b
encoding. We will also use an 8b/10b comma as a header just before the result pattern (this provides additional framing and marks the start of transmission. A different comma would fill any idle time between transmissions).
The 8b/10b codes as well as the 5-bit pattern labels provide error detection. Since the pattern labels have the same overhead as
8b/10b, we write the total number of bits for level 2 encoding as:
\begin{equation}
B_2 = 10 + N + K + 1.25(qK + C + \nu K)
\end{equation}
where $q$ is the number of label bits at level 0, $C$ is the number of context bits, and $\nu$ the number of cluster bits. 
The factor of 1.25 arises from replacing every 8 bits with 10 bits (or 4 label bits with 5 bits). In practice a few padding bits 
may need to be added on a transmission by transmission basis to make $(qK + C + \nu K)$ a multiple of 8. A different number of bins in the result pattern, if desired, can be accommodated by placing the additional binning bits in the 8b/10b encoded part. 

\begin{figure}[h]
\centerline{\includegraphics[width=0.65\textwidth]{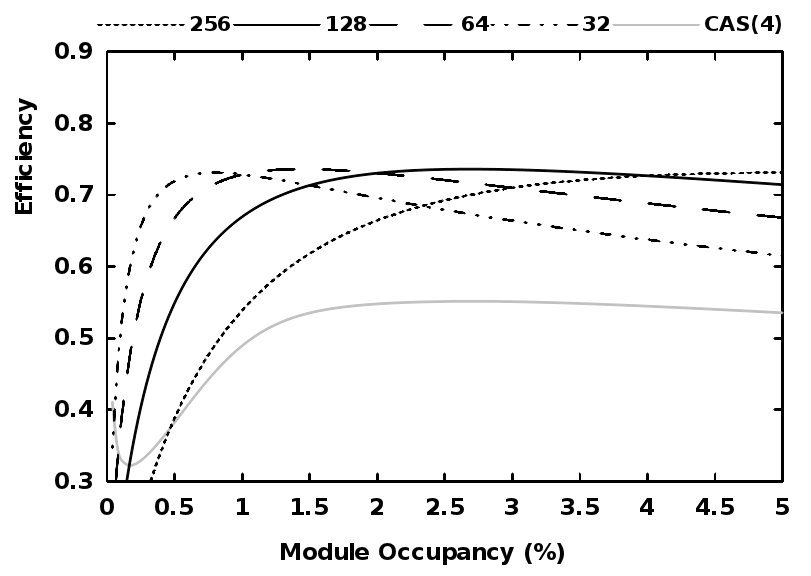}}
\caption{Level 2 efficiency, $\epsilon_2$, for POC encoding in a 10-chip module 
for different result pattern number of bins (256, 128, 64, 32),
as a function of module address occupancy. 
Also shown for comparison is $\epsilon_2$ for fixed packet CAS encoding with $n_p$=4.}
\label{fig:poc2}
\end{figure} 

\begin{figure}[h]
\centerline{\includegraphics[width=0.65\textwidth]{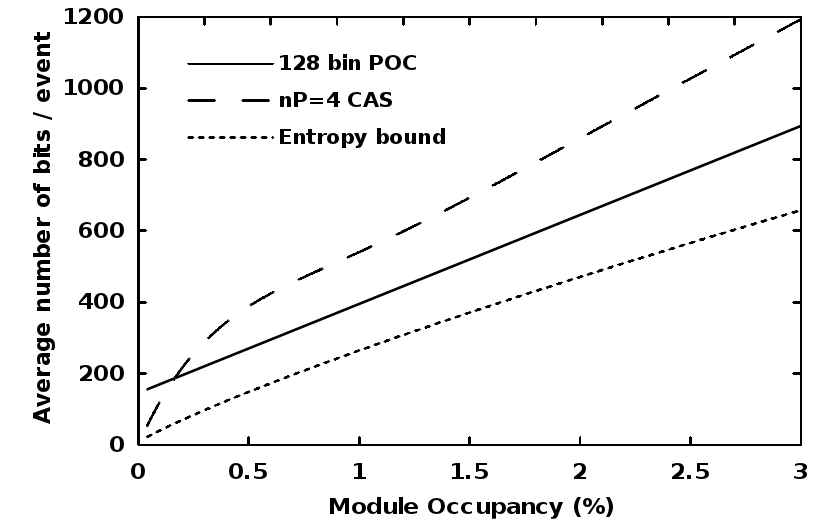}}
\caption{Average number of bits per event used by POC encoding in a 10-chip module 
for 128 result pattern bins, as a function of module occupancy,
compared to fixed packet CAS encoding with $n_P$=4. Also 
shown is the entropy bound for the minimum number of bits possible. 
All have engineering properties overhead and context data included.}
\label{fig:nbits}
\end{figure}

Fig.~\ref{fig:poc2} shows the level 2 efficiency for POC encoding applied to our 10-chip module case study,
compared to fixed packet CAS encoding. Recall that we made an optimistic assumption for the engineering properties
overhead added to fixed packet CAS, while the POC example is more conservative. 
POC encoding significantly outperforms fixed packet CAS even with this bias. 
For convenience of evaluating the practical 
impact on module operation, Fig.~\ref{fig:nbits} shows the average number of bits per event for the case of 128 bins 
in the POC result pattern, compared to fixed packet CAS with $n_P$=4, and to the entropy bound, 
all at level 2. A 0-3\% module occupancy range was used in this figure for clarity. 
The ratio of each of the top two curves to the entropy bound results in the corresponding efficiency curve in Fig.~\ref{fig:poc2}.

While we have not included it in the above calculation to keep it simple and conservative, further compression of multiple
pattern labels in the same bin is possible as described in Section~\ref{poc0}. 
It is instructive to see how that can be implemented within the DC-balanced scheme, because 
it also illustrates how the treatment of labels generalizes. 
For the first label in a bin there are always 10 possibilities in this particular 10-chip case, so a 5-bit symbol is used as described. 
For subsequent labels in the same bin, the symbols used depend on how many possibilities remain, assuming the labels are always written in increasing order. 
If only 6 possibilities remain, then a 4-bit label with 2 {\em ones} and 2 {\em zeroes} can be used. If only 3 possibilities remain, 
a 3-bit label with 2 {\em ones} and a single {\em zero} (or its inverse) can be used. While labels with an odd number of bits 
act to restore DC-balance as described, labels with an even number of bits have no effect (neither restore balance nor increase imbalance). 
Such labels will not compensate the imbalance introduced by the flag bits, and so cannot be used indiscriminately. 
In order to ensure that all transmissions will be balanced, regardless of $K$ or the distribution of the hits, an additional condition must be imposed when even-bit labels are to be used. This condition is that the number of remaining bins must be greater than the present imbalance (given by the DC-balance counter). Since bin delimiters act to restore balance, as long as enough bins are left one can guarantee that balance will be restored by the end of the result pattern. On the other hand, if use of an even-bit label would violate this condition, then the next larger odd-bit label must be used instead, which will act to restore DC-balance. This label selection method can now be used in general, not just for a 10-chip module, 
with the number of bits in the longest label as demanded by the number of source patterns being combined. 

\subsection{Pattern Overlay in the time domain and triggerless readout}

We have analyzed the combination of data from 10 chips in one module and in a single event. 
However, POC could also be applied entirely within a single chip to data from multiple events. 
In this case the pattern labels would correspond to an event counter rather
than identifying different physical chips. Combination in the time domain has the advantage that 
it does not need to match a physical hardware configuration, such as the number of chips in a module, and therefore the number 
of source patterns, $n_s$, can be optimized. However, in a triggered system, the combination of many source patterns each from 
a different triggered event would imply a significant latency. Thus POC encoding in the time domain is particularly suited 
to triggerless readout. For triggerless readout applications, achieving the maximum possible encoding efficiency will likely 
be a driving design consideration.      

\section{Conclusion and Outlook}

We have developed techniques to analyze the efficiency of data encoding and transmission for binary strip detector readout. 
This analysis has shown that in order to achieve high encoding efficiency for readout of low occupancy detectors,  
it is necessary to aggregate in a non-trivial way the data from either multiple readout chips or multiple frames. 
This is because the information entropy increases logarithmically with channel count, while the number of bits
used by any encoding method applied to a small set of channels (e.g. a chip) 
will increase linearly with the number of such sets. 
We have used this finding to develop an example encoding method called
Pattern Overlay Compression (POC), which indeed achieves high efficiency by aggregating data in 
a natural way, well suited to multi-chip modules or to combination of consecutive events in triggerless readout.  
Further work is in progress to develop a similar analysis of pixel detector readout, and to include signal
strength information, rather than just binary readout. 

\section{Appendix}

\subsection{Entropy of $k$ Random Addresses}

We wish to count the possible combinations of $k$ {\em ones} and $(n-k)$ {\em zeroes} 
with the condition that none of the {\em ones} are adjacent, as defined in Sec.~\ref{sec:compression}.
Obviously only patterns with $n \ge 2k-1$ can meet this condition.
We begin with the unique pattern 1010...101, containing $k$ {\em ones} and $(k-1)$ {\em zeroes}.
If $n = 2k-1$ this unique pattern is the only possibility. For $n > 2k-1$ 
(the case of interest in this paper) there will be a number of {\em zeroes} ``left over'' given by $n-(2k-1)$.
We now must count all the possible ways to insert these additional {\em zeroes}
into the 1010...101 pattern, in order to generate all 
possible $n$-bit patterns with exactly $k$ isolated {\em ones}. 
There are $(k+1)$ places where {\em zeroes} can be inserted 
(because the {\em zeroes} being inserted are indistinguishable from the {\em zeroes} already present). 
This is equivalent to the problem of chopping a string of $n-(2k-1)$ {\em zeroes} by inserting $k$ boundaries,
which can be represented by $k$ {\em ones}. The boundaries (or {\em ones}) can go anywhere. For example, if 
there are 4 {\em zeroes} to be chopped up by inserting 3 {\em ones}, we can have: 0000111, 0101010, 0110001, 1010010, etc.  
Thus, this is the familiar problem of choosing $k$ ones
out of $(n-k+1)$ bits, and the solution is the binomial coefficient $n-k+1 \choose k$. Therefore, the entropy of
all the random $n$-bit patterns with $k$ isolated {\em ones} is $log_{_2}$$n-k+1 \choose k$.

\subsection{CAS Efficiency as a Function of $n$ and Occupancy}

The number of bits needed by CAS is $B = k.log_2(n)$. Therefore,
\begin{eqnarray}
\epsilon_0 = & ln {n-k+1 \choose k} / [ k.ln(n)  ] \\
           = &  \{ ln[(n-k+1)!] - ln[(n-2k+1)!] - ln(k!) \}  / [ k.ln(n)  ]
\end{eqnarray}
We now substitute the following approximations valid for $n \gg k \gg 1$: 
$ln(x!)\approx x.ln(x) - x$ (Stirling's approximation) 
and $(n-k+1)[ln(n-k+1)-ln(n-2k+1)] \approx k$. After collecting terms we 
obtain 
\begin{equation}
\label{eq:casefficiency2}
\epsilon_0(n,k) \approx 1-{ln(k) - 1 \over ln(n)}
\end{equation}  
Eq.~\ref{eq:casefficiency} follows by substituting $ln(k) = ln(n) + ln(\alpha)$, where
$\alpha = k/n$.

\subsection{Entropy Overhead due to Engineering Properties}

We start from Eq.~\ref{eq:h2m} of section~\ref{EEP}. 
To obtain an expression for a continuous range of $H$ we 
use the approximation $2m \choose m$$ = 4^m/\sqrt{\pi m}$ and
then make the substitution $m \rightarrow x$, where $x$ is real instead of integer. 
Clearly this substitution is only possible in the approximate formula. We can 
now write,
\begin{eqnarray}
\label{eq:h2x}
H_{2x} &\approx & log_2( {4^x \over \sqrt{\pi x}}) \\
H_{2x} &\approx & 2x - \frac{_1}{^2}log_2(\pi x)
\end{eqnarray}
and 
\begin{equation}
\label{eq:E2x}
E(H_{2x}) = 2x - H_{2x} \approx \frac{_1}{^2}log_2(\pi x)
\end{equation}
We numerically plotted $E(H)$ vs. $H$ in Fig.~\ref{fig:hm} (dotted)
by looping over discrete values of $x$ and calculating the ordered pair $(H(x),E(x))$ for each $x$.
However, we would like an expression for $E(H)$ rather $E(x)$. 
From Eq.~\ref{eq:E2x} we have $x = 4^E/\pi$. Since we know $H \gg E(H)$, we can write,
\begin{equation}
2 \times 4^E/\pi \approx H + E \approx H
\end{equation}  
from which Eq.~\ref{eq:EH} follows. 
Eq.~\ref{eq:EH} is also shown in Fig.~\ref{fig:hm} (solid), where one can see that it is indeed a good approximation
to the numerical result.

\acknowledgments

This work was supported in part by the Office of High Energy Physics of the U.S. 
Department of Energy under contract DE-AC02-05CH11231. We thank A.~Grillo and J.~Agricola for helpful comments.


\begin{thebibliography}{99}

\bibitem{LHC}
L.~Evans and P.~Bryant (editors), ``The CERN Large Hadron Collider: Accelerator and Experiments: The LHC Machine,'' 
\jinst{3}{2008}{S08001}.


\bibitem{roi}
D.~Wardrope et al., \emph{Instrumentation of the upgraded ATLAS tracker with a double buffer 
front-end architecture for track triggering}, \jinst{7}{2012}{C08010}.

\bibitem{self1}
E.~Salvati et al.,
\emph{A Level-1 Track Trigger for CMS with double stack detectors and long barrel approach},
\jinst{7}{2012}{C08005}.

\bibitem{self2}
D.~Abbaneo et al.,
\emph{A hybrid module architecture for a prompt momentum discriminating tracker at HL-LHC},
\jinst{7}{2012}{C09001}.

\bibitem{self3}
A.~Sch\"oning et al.,
\emph{A Self Seeded First Level Track Trigger for ATLAS},
\jinst{7}{2012}{C10010}.




\bibitem{data-compression}
D.~Salomon, ``Data Compression,'' Springer-Verlag London Limited (2007).

\bibitem{zip}
PKWARE, inc., ``.ZIP File Format Specification,'' www.pkware.com (1989). 

\bibitem{entropy}
T.~M.~Cover and J.~A.~Thomas, ``Elements of Information Theory'' (2nd edition), John Wiley \& Sons, Hoboken, New Jersey (2006). 

\bibitem{8b10b}
A.~Widmer and P.~Franaszek,  ``A dc-balanced, partitioned-block, 8b/10b transmission
code,'' IBM Journal of Research and Development, 27(5), 440-451, (1983).


\bibitem{6466}
R.~C.~Walker and R.~Dugan, ``64b/66b low-overhead coding proposal for serial links,'' in IEEE 802.3 High Speed Study Group, update 1/12/00 (2000). 


\bibitem{hamming}
W.~W.~Peterson and D.~T.~Brown, ``Cyclic Codes for Error Detection,'' in Proceedings of the IRE, Volume 49,  Issue 1 (1961). 

\bibitem{sct}
F.~Campabadal et al., ``Design and performance of the ABCD3TA ASIC for readout of silicon
strip detectors in the ATLAS semiconductor tracker'', 
\emph{Nucl.\ Instrum.\ Meth.\ } {\bf A} 552,  p.292 (2005).



\end{thebibliography}
\end{document}